\documentstyle[preprint,aps]{revtex}

\begin{document}
\setcounter{page}{1}

\draft

\title{
 Remarks on the Hadronic Effect in Muon $g-2$: \\
 Low Energy Behavior of $ V^0 $-$\pi^+$ Scattering
}

\author{
 M. Hayakawa \footnotemark[1]
 }
\footnotetext[1]
{
 Electronic address: hayakawa@theory.kek.jp
}

\address{
 Department of Physics,
 National Laboratory for High Energy Physics (KEK), \\
 Tsukuba, Ibaraki, Japan 305\\
}

\preprint{\begin{minipage}{4cm}
           KEK-TH-491 \\
           hep-ph/9608225 \\
           August 1996 \\
          \end{minipage}
         }

\date{\today}

\maketitle

\begin{abstract}
 The behavior of the $V^0$-$\pi^+$ scattering amplitude
(where $V^0 = \rho^0 , \omega$ , or $\phi$)
in the low pion momentum limit
is studied motivated by its relevance to the theory
of muon $g - 2$.
 Current algebra analysis shows that its S-wave component
must vanish in the chiral limit
under general physical assumptions.
 We confirm this result
with various low energy models of vector meson.
 Our result justifies the calculation 
 of the charged pion loop part of
the hadronic light-by-light scattering contribution
to muon $ g-2 $
within the framework of hidden local symmetry model
of low energy hadron dynamics.
\end{abstract}


\narrowtext
%


%

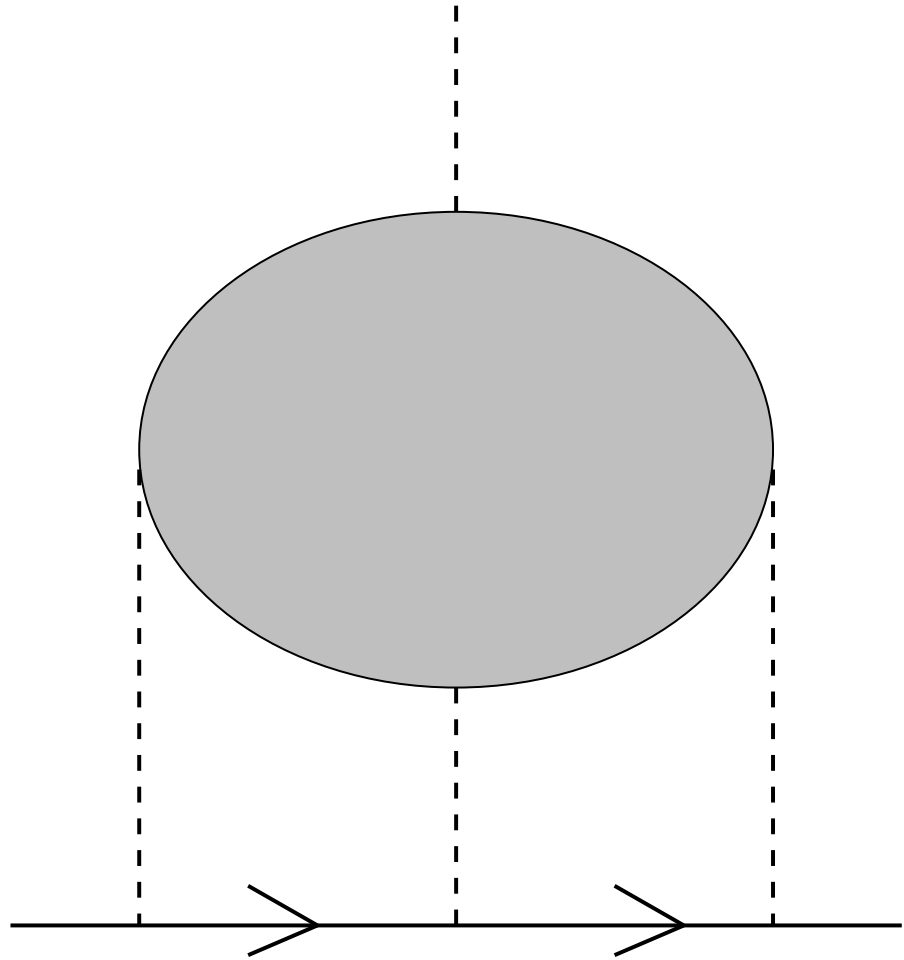
\begin{figure}
 \caption{Hadronic light-by-light scattering effect
          on muon $ g-2 $.
          The solid and dashed lines represent
          muon and photon respectively.
          The blob part corresponds to
          the light-by-light scattering amplitude by hadron(s).
          }
 \label{fig:light}
\end{figure}

 The ${\cal O}(\alpha^3) $ hadronic contribution
to the anomalous magnetic moment
of the muon contains
the hadronic light-by-light scattering effect
as shown in Fig. \ref{fig:light}.
 Evaluation of such a contribution
is currently restricted to purely theoretical analysis.
 Ref. \cite{K-had} examined this contribution
from the point of view of chiral dynamics.
 To answer some questions raised about this calculation
\cite{Einhorn},
it was recently re-analyzed
in Ref. \cite{Bijnens1,Bijnens2,HKS}.
 The study of Ref. \cite{HKS}
tried particularly to respond to the claim
that the naive vector meson dominance (VMD) model
used in Ref. \cite{K-had} does not maintain
the Ward identity inferred
from electromagnetic symmetry \cite{Einhorn}.
 Since this may affect the estimate
of charged pion loop contribution
to muon $g-2$,
Ref. \cite{HKS} attempted to rectify it
using the hidden local symmetry (HLS) approach \cite{Bando},
which incorporates the vector meson explicitly
in the effective low energy theory.
 It was found that the $ \rho^0 \rho^0 \pi^+ \pi^- $ vertex
is absent in the chiral limit in the HLS approach,
contradicting the naive VMD model.
 The resulting effective Lagrangian
enabled us to recover the relevant Ward identity
at the same time.

 However the first version of Ref. \cite{Bijnens2}
proposed a different effective Lagrangian
(eq. (5.6) in Ref. \cite{Bijnens2})
realizing VMD, which contains
the $ \rho^0 \rho^0 \pi^+ \pi^- $ vertex
in the chiral limit
justifying the naive VMD model of Ref. \cite{K-had}.
 In the final version of Ref. \cite{Bijnens2}
more complicated form of Lagrangian
has been written down
which consists of a few sets of
several infinite series of higher derivative terms
expanded in some mass scale $ M_X $.
 It gives a familiar VMD structure with this mass scale
for each the hadronic light-by-light scattering amplitude.
 The part
of it relevant for the muon anomaly
is that found in naive VMD model.
 Thus {\it when this scale is set to the rho-meson mass}
their numerical result of the charged pion contribution
to muon $g-2$
is actually close to the previous result \cite{K-had}
rather than that based on the HLS model \cite{HKS}.

 When we consider the $ \rho^0 $-$ \pi^+ $
scattering amplitude,
its leading behavior in low momentum limit
is quite different for theories without or with
the $ \rho^0 \rho^0 \pi^+ \pi^- $ vertex:
 The former vanishes at threshold while the latter does not.
 It would be reasonable
to expect that such a difference can be studied starting from
a chirally invariant Lagrangian
in view of
our experiences in the S-wave nucleon-pion scattering,
pion-pion scattering, and KSRF relation \cite{Kawarabayashi}.
In other words, at least
the exponent of momentum in the low momentum limit
would not depend on how the vector meson
is incorporated in the theory.
 As was mentioned above,
however, the effective Lagrangian of Ref. \cite{HKS} 
based on the hidden local symmetry approach
and that of Ref. \cite{Bijnens2}
based on the extended Nambu-Jona-Lasinio model
give different low energy behaviors.

 The purpose of this paper is
to settle this issue starting
from the chiral symmetry alone
without relying on any specific effective Lagrangian.
 Namely, we investigate the leading low energy behavior
of $ \rho^0 $-$ \pi^+ $ scattering
based on the traditional current algebra technique alone.
 Our analysis shows that its S-wave component
vanishes in the chiral limit.
 This implies that, when we consider
the effective theory of Goldstone boson and vector meson,
the $ \rho^0 \rho^0 \pi^+ \pi^- $ vertex
may first appear proportional to the current quark mass
or higher derivatives.

 In order to 
derive the low energy behavior of
$V^0$-$\pi^+$ ($ V^0 $ = $\rho^0$, $\omega$ or $\phi$ )
scattering amplitude 
let us start, 
as in the case of pion-nucleon scattering
\cite{Bando2},
from the algebraic identity
\begin{eqnarray}
 &&
  \displaystyle{
   (-i q^\mu) (i k^\nu)
   \int d^4 x e^{iq \cdot x} \int d^4 y e^{-ik \cdot y}
   \left<
    b\, {\bf P}_f\, \sigma
    \left|
     T J^+_{5\,\mu}(x)\, J^-_{5\,\nu}(y)
    \right|
    a\, {\bf P}_i\, \lambda
   \right>
  }
   \nonumber \\
 && =
 \displaystyle{
  \frac{1}{2} \int d^4 x e^{iq \cdot x}
   \int d^4 y e^{-ik \cdot y}
 } \nonumber \\
 && \quad \times
  \displaystyle{
   \left[
    (-iq^\mu)
     \left\{
       \delta(x^0 - y^0)
       \left< b\, {\bf P}_f\, \sigma
        \left|
         \left[
          J^-_{5\,0}({\bf y},x^0), J^+_{5\,\mu}
           ({\bf x},x^0)
         \right]
        \right| a\, {\bf P}_i\, \lambda
      \right>
     \right\}
   \right.
  } \nonumber \\
 && \quad \quad +
  \displaystyle{
    (ik^\nu)
     \left\{
       \delta(x^0 - y^0)
       \left< b\, {\bf P}_f\, \sigma
        \left|
         \left[
          J^+_{5\,0}({\bf x},x^0), J^-_{5\,\nu}
           ({\bf y},x^0)
         \right]
        \right| a\, {\bf P}_i\, \lambda
      \right>
     \right\}
  }
   \nonumber \\
 && \quad \quad
  \displaystyle{
    -iq^\mu
       \left< b\, {\bf P}_f\, \sigma
        \left| T
          J^+_{5\,\mu}(x)\, \partial^\nu J^-_{5\,\nu}(y)
        \right| a\, {\bf P}_i\, \lambda
     \right>
  } \nonumber \\
 && \quad \quad
  \displaystyle{
   \left.
     + ik^\nu
       \left< b\, {\bf P}_f\, \sigma
        \left| T
          \partial^\mu J^+_{5\,\mu}(x)\,
            J^-_{5\,\nu}(y)
        \right| a\, {\bf P}_i\, \lambda
      \right>
   \right]
  },
  \label{eq:identity}
\end{eqnarray}
where $ J^\pm_{5\mu} = \frac{1}{\sqrt{2}}
(J^1_{5\mu}  \pm i J^2_{5\mu} ) $ ,
$ \left| a\, {\bf P}_i\, \lambda \right> $
denotes the vector meson states of momentum $ {\bf P}_i $
(boldface letter means the usual three space momentum) 
and helicity
$ \lambda $.
 The internal label $ a $ signifies
the neutral vector meson states
in the nonet basis,
$ \rho^0 $, $ \omega $ and $ \phi $.
 Throughout this paper all the current quark masses
are set equal to zero.
 The masses of $ \omega $, $\phi$ and $ \rho^\pm $ are taken
equal for simplicity
although there is no a priori reason to expect it
even in the flavor symmetric limit.
 The divergence of the axial current,
$\partial^\mu  J^\pm_{5\,\mu}(x) $,
is kept in the last two terms of (\ref{eq:identity})
to accommodate 
possible influence of chiral anomaly.

 What we are interested in here is
whether the S-wave component of
$ V^0 $-$ \pi^+ $ scattering
amplitude is of order 1 or not
in the simultaneous limit of $ k^\mu, q^\mu \rightarrow 0 $.
 Since the pions are massless in the chiral-symmetric limit,
the above limit is restated more precisely 
as the $ \epsilon \rightarrow 0 $ limit where
\begin{eqnarray}
 q^\mu &=& ( \epsilon, \epsilon {\bf n}_q ),
  \nonumber \\
 k^\mu &=& ( \epsilon, \epsilon {\bf n}_k ),
  \quad \quad \left| {\bf n}_q \right|
    = \left| {\bf n}_k \right| = 1.
  \label{eq:pimomentum}
\end{eqnarray}
 Presumably the dominant contribution from the first two terms
on the RHS of Eq. (\ref{eq:identity}) is
\begin{eqnarray}
 &&
  \displaystyle{
   \frac{1}{2}
   \left[
    -iq^\mu
     \left< b\,{\bf P}_f\,\sigma
       \left|
         \int d^4 x
         \left[
           Q^-_5(x^0), J^+_{5\,\mu}({\bf x},x^0)
         \right]
       \right| a\,{\bf P}_i\,\lambda
     \right>
   \right.
  } \nonumber \\
 && \quad \ +
  \displaystyle{
   \left.
    ik^\nu
    \left< b\,{\bf P}_f\,\sigma
     \left|
      \int d^4 y
      \left[
        Q^+_5(x^0), J^-_{5\,\nu}({\bf y},y^0)
      \right]
     \right| a\,{\bf P}_i\,\lambda
    \right>
   \right]
  },
\end{eqnarray}
which arises from
the leading term of the expansion
$ e^{iq\cdot x} e^{-ik\cdot y}
 = 1 + iq\cdot x - ik\cdot y + \cdots $.
 The axial current $J^\pm_{5\,\mu}(x)$
transforms under the axial charge
$ Q^\mp_5 $ to the third component of
iso-vector current $ J^3_\mu(x) $:
\begin{eqnarray}
 \left[
  Q^-_5(x_0), J^+_{5\,\mu}({\bf x},x^0)
 \right]
 &=& - J^3_\mu({\bf x},x^0),
  \nonumber \\
 \left[
  Q^+_5(x_0), J^-_{5\,\mu}({\bf x},x^0)
 \right]
 &=& J^3_\mu({\bf x},x^0).
\end{eqnarray}
 Thus Eq. (\ref{eq:identity}) reduces, to the leading order
in $ \epsilon $,
to
\begin{eqnarray}
 &&
  \displaystyle{
   \lim_{\epsilon \rightarrow 0}
   \int d^4 x e^{iq\cdot x}
   \int d^4 y e^{-ik\cdot y}\,\,
   q^\mu k^\nu
   \left<
    b\,{\bf P}_f\,\sigma
    \left|
     T J^+_{5\,\mu}(x)\, J^-_{5\,\nu}(y)
    \right|
    a\,{\bf P}_i\,\lambda
   \right>
  } \nonumber \\
 && =
  \displaystyle{
   -\frac{1}{2} (q+k)^\mu \int d^4 x
    \left< b\,{\bf P}_f\,\sigma
     \left|
      J^3_\mu(x)
     \right| a\,{\bf P}_i\,\lambda
    \right>
  } \nonumber \\
 && \quad \ +
  \displaystyle{
   \frac{1}{2}
   \left[
    -iq^\mu
    \int d^4 x e^{iq \cdot x} \int d^4 y e^{-ik \cdot y}
       \left< b\, {\bf P}_f\, \sigma
        \left| T
          J^+_{5\,\mu}(x) \partial^\nu J^-_{5\,\nu}(y)
        \right| a\, {\bf P}_i\, \lambda
      \right>
   \right.
  } \nonumber \\
 && \quad \quad \quad \
  \displaystyle{
   \left.
     + ik^\nu
       \int d^4 x e^{iq \cdot x}
         \int d^4 y e^{-ik \cdot y}
       \left< b\, {\bf P}_f\, \sigma
        \left| T
          \partial^\mu J^+_{5\,\mu}(x) J^-_{5\,\nu}(y)
        \right| a\, {\bf P}_i\, \lambda
      \right>
   \right]
  }.
 \label{eq:relation}
\end{eqnarray}

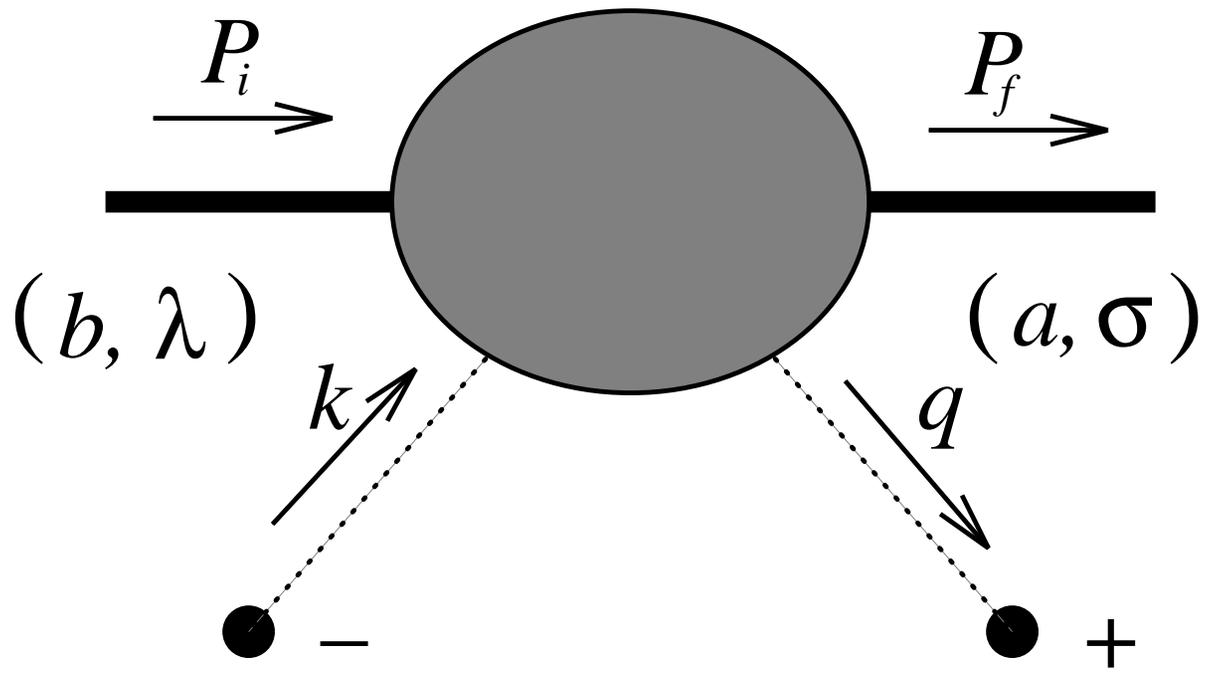
\begin{figure}
 \caption{$ V^0 $-$\pi^+$ scattering contribution.
          Here the dotted and bold lines denote the charged pion
          and the vector meson respectively.
          The blob part corresponds to the amplitude
          of pion and vector meson scattering.}
 \label{fig:category-a}
\end{figure}

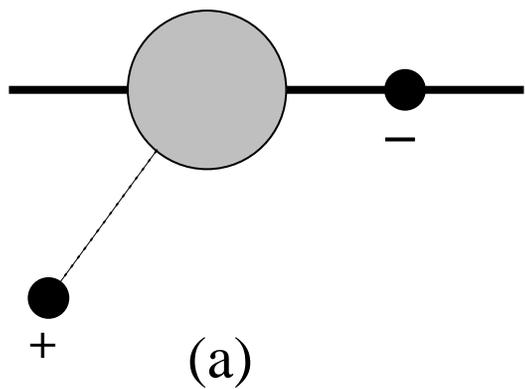
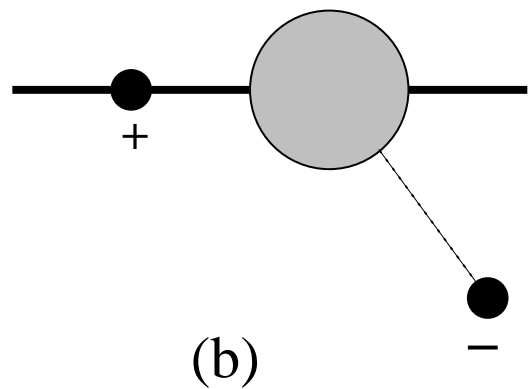
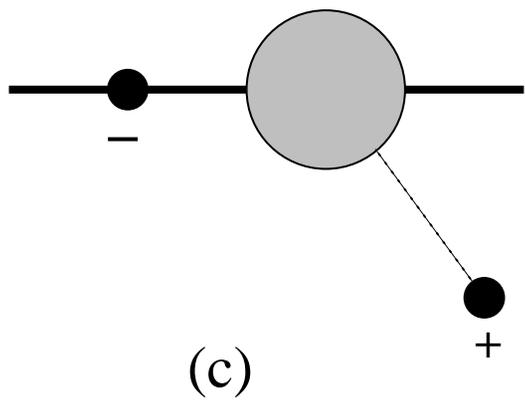
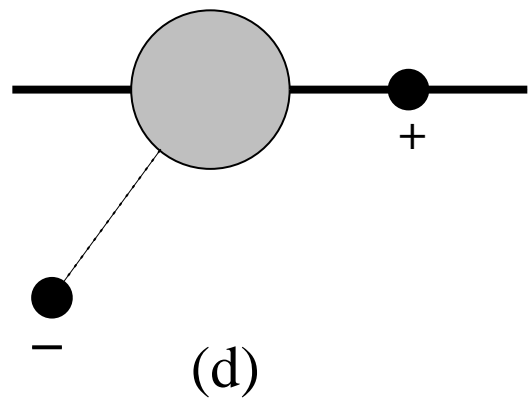
\begin{figure}
 \caption{ Singular components with one pion pole.
          The intermediate state here
          is the charged vector meson.
          The dark blob corresponds to
          the contact term of the axial current operator.
          The shaded blob denotes the anomalous $ V^0 V^- \pi^+ $
          ( or $ V^0 V^+ \pi^- $ ) vertex. }
 \label{fig:category-b}
\end{figure}

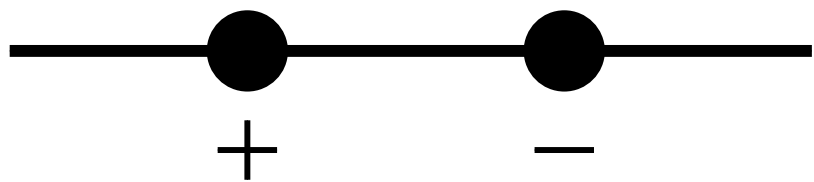
\begin{figure}
 \caption{ Singular components in which two axial currents
          are directly attached to the vector meson line.}
 \label{fig:category-c}
\end{figure}

 Let us first examine the LHS of eq. (\ref{eq:relation}).
 Recalling that the state vectors
describe neutral vector meson states,
the singular components with respect to $ q,\, k $ in
\begin{equation}
   \left<
    b\,{\bf P}_f\,\sigma
    \left|
     T J^+_{5\,\mu}(x)\, J^-_{5\,\nu}(y)
    \right|
    a\,{\bf P}_i\,\lambda
   \right> 
\end{equation}
can be classified into three categories
depicted
in Figs. \ref{fig:category-a}-\ref{fig:category-c},
analogous to the case of nucleon-pion scattering.
 The contribution of Fig. \ref{fig:category-a}
gives rise to the $ V^0 $-$\pi^+$ scattering amplitude
\begin{eqnarray}
{\cal A} ( {\rm Fig.}\, \ref{fig:category-a}) &=&
  - f_\pi^2 {\cal M}
    \left(
     V^a(P_i,\lambda) + \pi^+(k)
      \rightarrow V^b(P_f,\sigma) + \pi^+(q)
    \right) \nonumber \\
 && \quad \times
  i (2\pi)^4 \delta^4(P_f + q - P_i - k).
\end{eqnarray}
 The diagrams
of Figs. \ref{fig:category-b} and \ref{fig:category-c}
are singular if and only if the intermediate states
are degenerate in mass
with the external vector meson.
 Recalling the realistic hadronic spectrum,
we may assume that the non-anomalous parts
of the current operator in these diagrams do not induce a singular
contribution:
For instance, axial-vector meson (such as $ A_1 $)
is much heavier than $\rho , \omega $, or $\phi$.
 Thus the only possible source of singular contribution 
is the anomalous part of the currents with
the propagation of the charged $\rho$ meson
as the intermediate state.
 Isospin invariance (or conservation of G-parity)
guarantees that this contribution
vanishes for the $ \rho^0 $-$ \pi^+ $ scattering.
 In the other cases,
each diagram multiplied by a factor $ q^\mu\, k^\nu $
may give contribution of order $ \epsilon $ at most 
to the LHS of Eq. (\ref{eq:relation}).
(See eq. (\ref{eq:pimomentum}) for the definition of $ \epsilon $.)
 Actually the S-wave component of 
the sum of diagrams in Figs. \ref{fig:category-b}(a)
and \ref{fig:category-b}(d),
Figs. \ref{fig:category-b}(b) and \ref{fig:category-b}(c),
are found to be of order $ \epsilon^2 $ respectively.
 The demonstration is quite similar as in the case of
pion-nucleon scattering.
 Likewise one can see that the cancellation
of order one terms occurs
between the diagrams in Fig. \ref{fig:category-c} .

 Similar consideration persists as well for
the last two terms on the RHS of Eq. (\ref{eq:relation}),
due to the absence of Goldstone-pole contribution
in this case (no such a pole in $ \partial^\mu J_{5\,\mu}^\pm $),
and they vanish in the limit $ \epsilon \rightarrow 0 $.

 Finally, since
the vector current $J_\mu^3$ transforms
as $ {\cal C} J^3_\mu {\cal C}^\dagger = - J^3_\mu $
under charge conjugation,
the first term of the RHS,
which is the matrix element of $J_\mu^3$
between two $V^0$ states,
vanishes identically.

 Thus we conclude that
the $ V^0 $-$ \pi^+ $ scattering amplitude
must vanish in the limit of $ \epsilon \rightarrow 0 $.

%

 Various models of vector meson proposed thus far
will help to check to the general argument
\cite{Birse}.

 Since the case of HLS approach has been explored
before \cite{HKS},
we do not repeat it again.

\subsection{Massive Vector Yang-Mills Approach}
\label{yang-mills}

 First we take up the massive Yang-Mills approach
incorporating vector meson only
\cite{Weinberg,Kaymakcalan}.
 Following the procedure of Ref. \cite{Kaymakcalan},
the vector meson of nonet $ V_\mu $ is introduced
from $ A^L_\mu $ and $ A^R_\mu $
which transform under the local chiral transformation
$ (g_L,g_R) \in G \equiv U(3)_L \times U(3)_R $
\begin{eqnarray}
 A^{L\,\prime}_\mu &=&
  g_L A^L_\mu g_L^\dagger
   + \frac{i}{g} g_L \partial_\mu g_L^\dagger,
  \nonumber \\
 A^{R\,\prime}_\mu &=&
  g_R A^R_\mu g_R^\dagger
   + \frac{i}{g} g_R \partial_\mu g_R^\dagger,
\end{eqnarray}
in such a way that
\begin{eqnarray}
 A^L_\mu &=&
  \xi(\pi) \rho_\mu \xi(\pi)^\dagger
    + \frac{i}{g} \xi(\pi) \partial_\mu \xi(\pi)^\dagger,
    \nonumber \\  
 A^R_\mu &=&
  \xi(\pi)^\dagger \rho_\mu \xi(\pi)
    + \frac{i}{g} \xi(\pi)^\dagger \partial_\mu \xi(\pi),
 \label{eq:rho-A}
\end{eqnarray}
with the Goldstone boson matrix
\begin{eqnarray}
 \xi(\pi) &=&
  \displaystyle{
    {\rm exp}
     \left(
      i \frac{\pi}{f_\pi}
     \right)
  }, \nonumber \\
 \pi &=&
   \displaystyle{
    \pi^a T^a
   } .
\end{eqnarray}
( The group generator $ T^a $ is normalized as
$ 2\,{\rm Tr}(T^a\,T^b) = \delta^{ab} $
throughout the paper. )
 The transformation law of $ V_\mu $
is non-linear \cite{Weinberg}
\begin{equation}
 V^\prime_\mu
   = h(\pi,g_L,g_R) V_\mu h(\pi,g_L,g_R)^\dagger
     + h(\pi,g_L,g_R) \partial_\mu h(\pi,g_L,g_R)^\dagger,       
\end{equation}
as is verified from
\begin{equation}
 \xi(\pi^\prime) = g_L \xi(\pi) h(\pi,g_L,g_R)^\dagger
                 = h(\pi,g_L,g_R) \xi(\pi) g_R^\dagger,
 \label{eq:h}
\end{equation}
where $ h(\pi,g_L,g_R) $ is an element of the vector group
$ {\rm U(3)_V} $.
 If we introduce the external gauge fields $ {\cal R}_\mu $,
$ {\cal L}_\mu $ which transform as
\begin{eqnarray}
 {\cal L}^\prime_\mu &=&
  g_L {\cal L}_\mu g_L^\dagger
   + i g_L \partial_\mu g_L^\dagger,
  \nonumber \\
 {\cal R}^\prime_\mu &=&
  g_R {\cal R}_\mu g_R^\dagger
   + i g_R \partial_\mu g_R^\dagger,
\end{eqnarray}
the local $ {\rm U(3)_L \times U(3)_R }$ invariant terms
bilinear at most in $ V_\mu $ with no further derivatives
are
\begin{eqnarray}
 &&
 \displaystyle{
  m_0^2\, {\rm Tr}
  \left[
   (A^R_\mu - \frac{1}{g} {\cal R}_\mu)^2
   + (A^L_\mu - \frac{1}{g} {\cal L}_\mu)^2
  \right]
  - B\, {\rm Tr}
     \left[
      (A^L_\mu - \frac{1}{g} {\cal L}_\mu) U
      (A^{R\,\mu} - \frac{1}{g} {\cal R}^\mu) U^\dagger
     \right],
 }
 \label{eq:rho-2}
\end{eqnarray}
where $ U \equiv \xi^2 $, plus the Yang-Mills term.
(In general the terms like
\begin{equation}
 \left\{
  {\rm Tr}(A^R_\mu - \frac{1}{g} {\cal R}_\mu)
  \right\}^2
+ \left\{
   {\rm Tr}(A^L_\mu - \frac{1}{g} {\cal L}_\mu)
  \right\}^2 ,
\end{equation}
are also permitted.
 However the conclusion will not be affected by the presence of
those terms
so that these kinds of terms will not be mentioned explicitly.
 The same remark should also be recalled hereafter.)
 From eq. (\ref{eq:rho-A}), eq. (\ref{eq:rho-2})
contains two $ V^\prime$ s only in the form
of the vector meson mass term.
 This is the same situation as in the HLS approach;
the $ V^0 V^0 \pi^+ \pi^- $ vertex appears
with explicit symmetry violation by quark mass
or with higher derivatives, e.g.,
\begin{eqnarray}
 &&
 \displaystyle{
  {\rm Tr}\,
   \left[
    \left(
     A^L_\mu - \frac{1}{g} {\cal L}_\mu
    \right) {\cal M}
    \left(
     A^{R\,\mu} - \frac{1}{g} {\cal R}^\mu
    \right)
    U
   \right]
  } \nonumber \\
 && +
 \displaystyle{
  {\rm Tr}\,
   \left[
    \left(
     A^L_\mu - \frac{1}{g} {\cal L}_\mu
    \right) U
    \left(
     A^{R\,\mu} - \frac{1}{g} {\cal R}^\mu
    \right)
    {\cal M}^\dagger
   \right]
  },
 \label{eq:higher}
\end{eqnarray}
where the current quark mass matrix $ {\cal M} $ is
considered to transform as
\begin{equation}
 {\cal M} \rightarrow g_L {\cal M} g_R^\dagger,
\end{equation}
when the quark mass term appears
as $ -\bar{q}_L {\cal M} q_R $ in QCD Langrangian.
( Hereafter those terms as in (\ref{eq:higher})
will be called as higher-order terms in derivative expansion.)

\subsection{Generalized Hidden Local Symmetry Approach}
\label{sec:generalized-HLS}

 The generalized HLS approach incorporates
the vector and axial-vector mesons as the gauge fields of
HLS
$ {\rm[ U(3)_L\times U(3)_R]_{local}} $ \cite{Bando}.
 Since its description is rather cumbersome,
the reader should consult with Ref. \cite{Bando}.
 Eqs. (7.63) and (7.64) of Ref. \cite{Bando} are
the terms which involve at most two vector
or axial-vector mesons of lowest order in derivatives.
 In the unitary gauge of HLS, these become
\begin{eqnarray}
 {\cal L} &=&
  \displaystyle{
   a f_\pi^2\,{\rm Tr}
   \left[
    \left(
     g V_\mu - \frac{i}{2f_\pi^2}
               [\pi,\partial_\mu \pi]
     + \cdots
    \right)^2
   \right]
  } \nonumber \\
 && +
  \displaystyle{
   (b+c) f_\pi^2\,{\rm Tr}
   \left[
    \left(
     g A_\mu - \frac{b}{b+c} \partial_\mu \pi
      + \cdots
    \right)^2
   \right]
   + {\rm kinetic\ terms}
  },
  \label{eq:g-HLS}
\end{eqnarray}
where
$ V_\mu $ and $ A_\mu $ are the vector and axial-vector nonets,
and $ a, b $ and $ c $ are the constants to be determined from
the experiments.
 The axial hidden gauge transformation
\begin{eqnarray}
 (V_\mu + A_\mu) &\rightarrow&
  \displaystyle{
   (V^\prime_\mu + A^\prime_\mu) =
    \frac{i}{g}\,\eta\,\partial_\mu \eta^\dagger
    + \eta\,(V_\mu + A_\mu)\,\eta^\dagger
  }, \nonumber \\
 (V_\mu - A_\mu) &\rightarrow&
  \displaystyle{
   (V^\prime_\mu - A^\prime_\mu) =
    \frac{i}{g}\,\eta^\dagger\,\partial_\mu \eta
    + \eta^\dagger\,(V_\mu - A_\mu)\,\eta
  },
\end{eqnarray}
where
\begin{equation}
 \eta = {\rm exp}\,
  \left(
   i \frac{b}{b+c} \frac{\pi}{f_\pi}
  \right) ,
\end{equation}
rotates away the transition term
between $ A_\mu $ and $ \pi $ in eq.(\ref{eq:g-HLS}).
 This manipulation induces the $ \rho^0 \rho^0 \pi^+ \pi^- $
interaction 
\begin{equation}
 \frac{1}{f_\pi^2} (M_{A_1}^2 - M_V^2)
  \left(\frac{b}{b+c}\right)^2
   \rho^0_\mu \rho^{0\,\mu} \pi^+ \pi^- ,
\end{equation}
with $ M_{A_1}^2 = (b+c) f_\pi^2 g^2 $
and $ M_V^2 = a f_\pi^2 g^2 $,
to the Lagrangian.
 However the $ \pi^+ \rho^0 A_1^- $ interaction also appears
at the same time;
\begin{equation}
 \frac{i}{f_\pi^2} (M_{A_1}^2 - M_V^2)\,
  \frac{b}{b+c}\,
  \left(
   A_1^{-\,\mu} \rho^0_\mu \pi^+
   - A_1^{+\,\mu} \rho^0_\mu \pi^-
  \right) .
\end{equation}
 As a result the invariant amplitude of $ \rho^0 $-$  \pi^+ $
scattering becomes
\begin{eqnarray}
 &&
 \displaystyle{
  {\cal M}\left(
           \rho^0(P_i,\lambda) + \pi^+(k) \rightarrow
           \rho^0(P_f,\sigma)  + \pi^+(q)
          \right)
 } \nonumber \\
 && =
  \displaystyle{
   \varepsilon^{(\sigma)\,*}_\mu({\bf P}_f)
   \varepsilon^{(\lambda)}_\nu({\bf P}_i)
   \left[
    \frac{2(M_{A_1}^2 - M_V^2)}{f_\pi^2}
     \left( \frac{b}{b+c} \right)^2 g_{\mu\nu}
   \right.
  } \nonumber \\
 && \quad +
  \displaystyle{
   \frac{(M_{A_1}^2 - M_V^2)^2}{f_\pi^2}
   \left( \frac{b}{b+c} \right)^2
    \left\{
     \frac{1}{(P_i + k)^2 - M_{A_1}^2}\,
     \left(
      g^{\mu\nu} - \frac{(P_i + k)^\mu (P_i + k)^\nu}{M_{A_1}^2}
     \right)
    \right.
  } \nonumber \\
 && \qquad \qquad \qquad \qquad \qquad \qquad \quad +
  \displaystyle{
    \left.
     \left.
     \frac{1}{(P_f - k)^2 - M_{A_1}^2}\,
     \left(
      g^{\mu\nu} - \frac{(P_f - k)^\mu (P_f - k)^\nu}{M_{A_1}^2}
     \right)
    \right\}
    \right]
  }. \nonumber \\
\end{eqnarray}
 It is an easy algebraic task
to observe that the S-wave component of this amplitude
is actually
${\cal O}(\epsilon^2) $ in the limit (\ref{eq:pimomentum})
, consistent with the low energy theorem
shown in the above general discussion.

\subsection{Antisymmetric Tensor Field Approach}
\label{sec:antisymmetric}

 The vector and axial-vector mesons can be
incorporated into the effective theory
as antisymmetric tensor fields \cite{Ecker}.
 Then there does not arise
a mixing between axial-vector mesons and Goldstone bosons.

 The nonet $ R_{\mu\nu} $ ( $ R = V$, or $ A $ ) transforms
under $ G $ as
\begin{equation}
 R^\prime_{\mu\nu} =
  h(\pi,g_L,g_R) R_{\mu\nu} h(\pi,g_L,g_R)^\dagger,
\end{equation}
by $ h(\pi,g_L,g_R) $ defined in eq. (\ref{eq:h}).
 The possible $ R^2 $ terms with the least numbers of derivatives
and quark mass which include pion fields are, e.g.,
\begin{eqnarray}
 &&
  \displaystyle{
   {\rm Tr}\,
    \left[
     \left(
      \xi^\dagger {\cal M} \xi^\dagger
      + \xi {\cal M}^\dagger \xi
     \right)
     R_{\mu\nu} R^{\mu\nu}
    \right]
  }, \nonumber \\
 &&
  \displaystyle{
   {\rm Tr}\,
    \left[
     R_{\mu\alpha} u^\alpha R_{\mu\beta} u_\beta
    \right]
  }, \nonumber \\
 &&
  \displaystyle{
   {\rm Tr}\,
   \left[
    A_{\mu\nu}
    \left[
     \left(
      \xi^\dagger {\cal M} \xi^\dagger
       - \xi {\cal M}^\dagger \xi
     \right),
     V^{\mu\nu}
    \right]
   \right]
  },
\end{eqnarray}
where
\begin{eqnarray}
 &&
  u_\mu = i \xi^\dagger (D_\mu U) \xi^\dagger, \nonumber \\
 &&
  D_\mu U = \partial_\mu U + iU{\cal R}_\mu
            - i {\cal L}_\mu U.
\end{eqnarray}
 Contrary to the generalized HLS approach,
both $ \rho^0 \rho^0 \pi^+ \pi^- $ and $ \pi^+ \rho^0 a_1^- $
vertices do not exist in the chiral limit.
 Such a respect is similar to the one seen
in the massive Yang-Mills approach
to incorporate vector and axial-vector mesons.
(See eq.(7.105) of Ref. \cite{Bando}.)


 To summarize,
current algebra analysis predicts that
the amplitude of $ \rho^0 $-$\pi^+ $ scattering shall vanish
in the low pion momentum.
 We have checked this fact based on the various
chiral-symmetric models of vector meson.
 The effective Lagrangian of Ref. \cite{Bijnens2}
are not written to include the vector meson field explicitly.
 Thus the problem is whether the structure found in
their formula
for the hadronic light-by-light scattering amplitude
can be obtained starting
from some chiral-symmetric model of vector meson.
 The result here seems to indicate
that it is impossible in the chiral limit.

 The effective Lagrangian of Ref. \cite{Bijnens2}
may be used to describe
the effects to be induced by integrating out
some resonance (other than rho meson)
if it scatters with pion of order one in the low energy limit.
 But setting $M_X$ to the mass ( $\ge$ 1 [GeV] )
of such a resonance will show
that the corresponding contribution to
the muon $ g-2 $
is negligible compared to
the ``rho-meson'' contribution
based on the HLS Lagrangian \cite{HKS},
a chiral-symmetric model of the vector meson.

\acknowledgements
 The author thanks K. Kinoshita
for his careful reading the manuscript.
 Valuable discussions with M. Harada, A. I. Sanda
and K. Yamawaki are also gratefully acknowledged.
 This research is partially supported by
Japan Society for the Promotion of Science
for Japanese Junior Scientists.

%

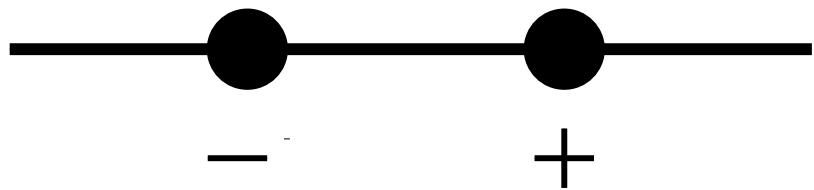
\end{document}